\documentclass[letter,bibyear]{aa} 

\usepackage{epsfig}
\usepackage{amsmath}
\usepackage{subfigure}
\usepackage{natbib}
\usepackage{multicol}
\usepackage{multirow}
\usepackage{color}
\usepackage{float}
\usepackage{longtable}
\usepackage{graphicx}
\usepackage{epstopdf}
\usepackage{booktabs}
\usepackage{txfonts}

\newcommand{\jms}{J.~Mol.~Spectrosc.}   
\newcommand{\jmst}{J.~Mol.~Struct.}   

\newcommand{\kms}{km s$^{-1}$}

\begin{document}

\title{Discovery of five cyano derivatives of propene with the 
QUIJOTE$^1$ line survey
\thanks{Based on observations carried out
with the Yebes 40m telescope (projects 19A003,
20A014, 20D023, and 21A011). The 40m
radiotelescope at Yebes Observatory is operated by the Spanish Geographic 
Institute
(IGN, Ministerio de Transportes, Movilidad y Agenda Urbana).}}

\author{
J.~Cernicharo\inst{1},
R.~Fuentetaja\inst{1},
C.~Cabezas\inst{1},
M.~Ag\'undez\inst{1},
N.~Marcelino\inst{2},
B.~Tercero\inst{2,3},
J.~R.~Pardo\inst{1}, and
P.~de~Vicente\inst{3}
}

\institute{Grupo de Astrof\'isica Molecular, Instituto de F\'isica Fundamental (IFF-CSIC),
C/ Serrano 121, 28006 Madrid, Spain\\ \email jose.cernicharo@csic.es
\and Centro de Desarrollos Tecnol\'ogicos, Observatorio de Yebes (IGN), 19141 Yebes, Guadalajara, Spain
\and Observatorio Astron\'omico Nacional (OAN, IGN), Madrid, Spain
}

\date{Received; accepted}

\abstract{
We report the discovery of five cyano derivatives of propene
towards \mbox{TMC-1} with the QUIJOTE$^1$ line survey: $trans$ and $cis$-crotononitrile ($t$-CH$_3$CHCHCN, $c$-CH$_3$CHCHCN), 
methacrylonitrile (CH$_2$C(CH$_3$)CN),
and $gauche$ and $cis$-allyl cyanide ($g$-CH$_2$CHCH$_2$CN and $c$-CH$_2$CHCH$_2$CN).
The observed transitions allowed us to derive a common rotational temperature of 7$\pm$1 K for all them. The
derived column densities are N($t$-CH$_3$CHCHCN)=(5$\pm$0.5)$\times$10$^{10}$ cm$^{-2}$, 
N($c$-CH$_3$CHCHCN)=(1.3$\pm$0.2)$\times$10$^{11}$ cm$^{-2}$, 
N(CH$_2$C(CH$_3$)CN)=(1.0$\pm$0.1)$\times$10$^{11}$ cm$^{-2}$, 
N($g$-CH$_2$CHCH$_2$CN)=(8.0$\pm$0.8)$\times$10$^{10}$ cm$^{-2}$, 
and
N($c$-CH$_2$CHCH$_2$CN)=(7.0$\pm$0.7)$\times$10$^{10}$ cm$^{-2}$,
respectively. The abundance of cyano-propene relative to that of
propene is thus $\sim$10$^{-2}$, which is considerably lower than those of other cyano
derivatives of abundant hydrocarbons. Upper limits are obtained for two ethynyl derivatives
of propene ($E$ and $Z$-CH$_3$CHCHCCH).

}

\titlerunning{Cyano-propene in TMC-1}
\authorrunning{Cernicharo et al.}

\maketitle

\section{Introduction}
Propene (also called propylene, CH$_2$CHCH$_3$)
is the most saturated hydrocarbon ever detected in space through radio
astronomical techniques. In spite of its weak dipole moment, several doublets of its $A$ and $E$ species
were observed by \citet{Marcelino2007} towards \mbox{TMC-1} with the IRAM 30m radio telescope. They 
derived, surprisingly, a very large column density, 4$\times$10$^{13}$ cm$^{-2}$. 
The molecule has also been detected towards four other cold dense clouds, Lupus-1\,A, L1495B,
L1521F, and Serpens South \citep{Agundez2015}, with column densities similar to those of \mbox{TMC-1 within a factor
of two}. More recently, this species has also been detected towards
the hot corino IRAS 16293-2422B \citep{Manigand2021}.

Since its discovery, the chemical paths leading to the formation of propene have been subject to debate.
\citet{Marcelino2007} proposed that the formation of propene could occur through reactions of radiative associations between 
hydrocarbon cations and H$_2$. However, laboratory
experiments and detailed calculations by \citet{Lin2013} have shown that these reactions are
not efficient at the low temperatures of 10\,K prevailing in TMC-1. 
A recent analysis of the gas-phase reactions producing propene has been provided by \citet{Hickson2016} who 
conclude
that it cannot be synthesized in the gas phase and suggest that it is formed by hydrogenation
of C$_3$ on the surface of dust grains followed by non-thermal desorption processes. 
In the hot corino IRAS 16293-2422B, models including hydrogenation and radical-radical additions on
grain surfaces seem to explain the
observed abundance of propene and other species \citep{Manigand2021}. However, the kinetic temperature
of the dust and gas in this source is much higher than in \mbox{TMC-1}.

The QUIJOTE\footnote{\textbf{Q}-band \textbf{U}ltrasensitive \textbf{I}nspection \textbf{J}ourney 
to the \textbf{O}bscure \textbf{T}MC-1 \textbf{E}nvironment} 
line survey of \mbox{TMC-1} \citep{Cernicharo2021a} performed with the Yebes 40m radio telescope has
permitted 29 new molecular species to be detected in the last months, among them are cyclic hydrocarbons
such as indene, benzyne, and 
cyclopentadiene (see, e.g. \citealt{Cernicharo2021a,Cernicharo2021b,Cernicharo2021c,Cernicharo2021d} and references therein). 
Propargyl, CH$_2$CCH, has been found to be one of the most abundant hydrocarbon radicals in this source
\citep{Agundez2021,Agundez2022a}. Other hydrocarbons
such as CH$_2$CCHCCH and CH$_2$CHCCH have been found to have large abundances as well \citep{Cernicharo2021b,Cernicharo2021d}. 
The discovery of this variety of hydrocarbons, with different degrees of saturation and sizes, strongly suggests that large 
and complex hydrocarbons 
such as polycyclic aromatic hydrocarbons are formed through a bottom-up mechanism involving small hydrocarbons as 
intermediate species. We still lack a correct 
picture of the complex hydrocarbon chemistry at work in cold dark clouds. Discovering new hydrocarbons and their
CN and CCH derivatives could help to elucidate the origin of these species, either gas-phase, grain-surface, or both.

\begin{figure}
\centering
\includegraphics[width=0.4\textwidth]{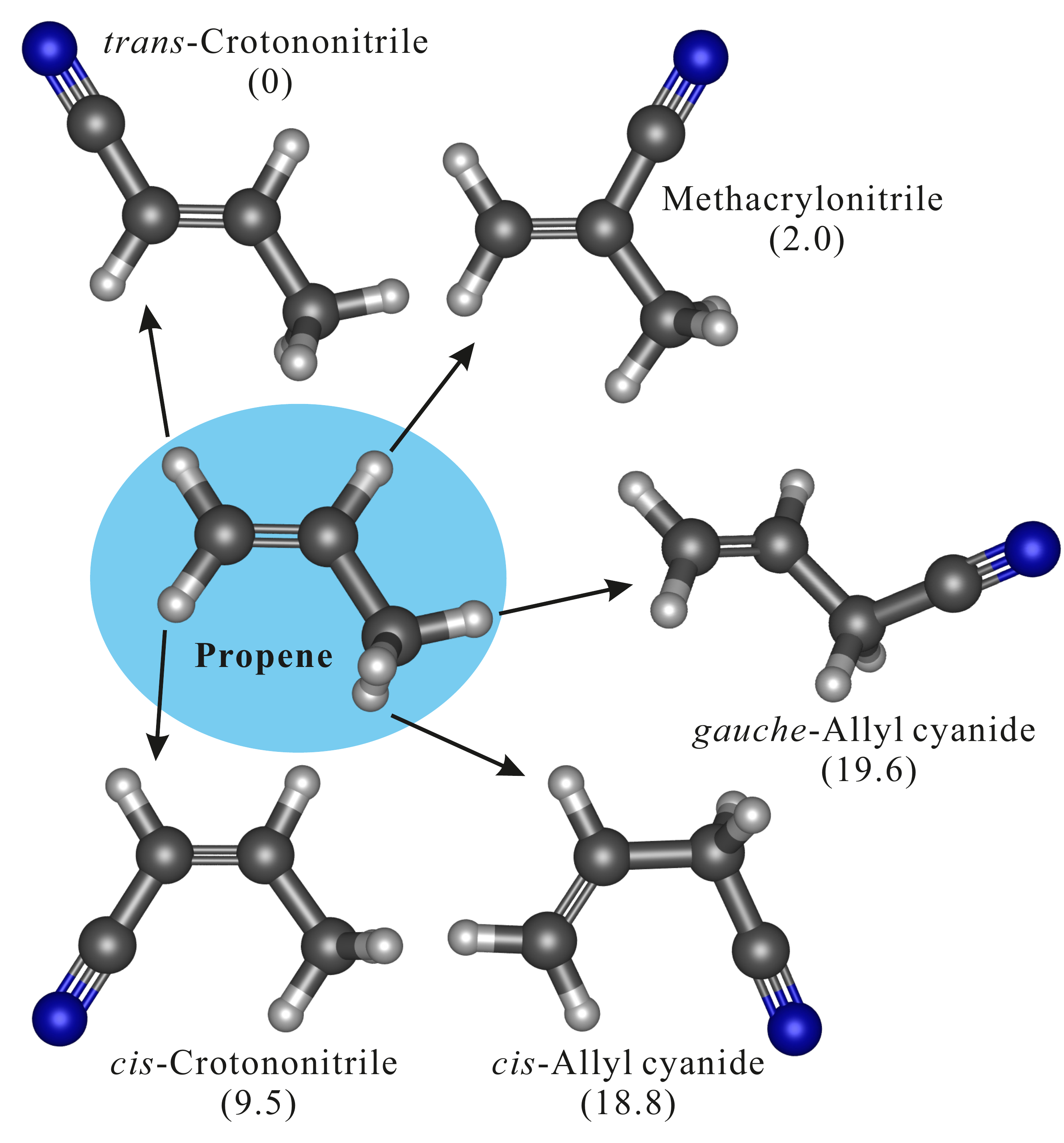}
\caption{Structures of the five cyano derivatives of propene (CH$_3$CHCH$_2$). The relative
energies in kJ mol$^{-1}$ are given between parentheses. The arrows indicate the H atoms substituted by
the CN group.}
\label{fig_structures}
\end{figure}

In this Letter, we report the discovery of five cyano derivatives of propene (CH$_2$CHCH$_3$). 
Their structures are shown in Fig. \ref{fig_structures}.
The detected species are $trans$ and $cis$-crotononitrile ($t$-CH$_3$CHCHCN, $c$-CH$_3$CHCHCN), methacrylonitrile (CH$_2$C(CH$_3$)CN), and 
$gauche$ and $cis$-allyl cyanide ($g$-CH$_2$CHCH$_2$CN, $c$-CH$_2$CHCH$_2$CN). 

\section{Observations} \label{observations}

New receivers, built within the Nanocosmos project\footnote{\texttt{https://nanocosmos.iff.csic.es/}},
and installed at the Yebes 40m radiotelescope, were used
for the observations of \mbox{TMC-1}
($\alpha_{J2000}=4^{\rm h} 41^{\rm  m} 41.9^{\rm s}$ and $\delta_{J2000}=
+25^\circ 41' 27.0''$). A detailed description of the system is 
given by \citet{Tercero2021}. Details of the QUIJOTE line survey are provided by \citet{Cernicharo2021a}.
The observations were carried out during different observing runs between November 2019 and January 2022.
The receiver consists of two cold high electron mobility transistor amplifiers covering the
31.0-50.4 GHz band with horizontal and vertical             
polarizations. Receiver temperatures in the runs achieved during 2020 vary from 22 K at 32 GHz
to 42 K at 50 GHz. Some power adaptation in the down-conversion chains have reduced
the receiver temperatures during 2021 to 16\,K at 32 GHz and 25\,K at 50 GHz.
The backends are $2\times8\times2.5$ GHz fast Fourier transform spectrometers
with a spectral resolution of 38.15 kHz,
providing the whole coverage of the Q-band in both polarizations.

The data presented here
correspond to observations carried out in the period 2019-2021 and correspond to 427 hours of observing time 
on the source, of which 254 and 173 hours were acquired with a 
frequency switching throw of 10 MHz and 8 MHz, respectively. 
The intensity scale used in this work, the antenna temperature
($T_A^*$), was calibrated using two absorbers at different temperatures and the
atmospheric transmission model (ATM; \citealt{Cernicharo1985, Pardo2001}).
The antenna temperature has an estimated uncertainty of 10~\% and can be 
converted to main beam brightness temperature, $T_{mb}$, by dividing by $B_{\rm eff}$/$F_{\rm eff}$. For the Yebes 40m 
telescope, $B_{\rm eff}$\,=\,0.738\,$\exp$[$-$($\nu$(GHz)/72.2)$^2$] and $F_{\rm eff}$\,=\,0.97 \citep{Tercero2021}.
All data were analysed using the GILDAS package\footnote{\texttt{http://www.iram.fr/IRAMFR/GILDAS}}.

\begin{figure*}
\centering
\includegraphics[width=0.99\textwidth]{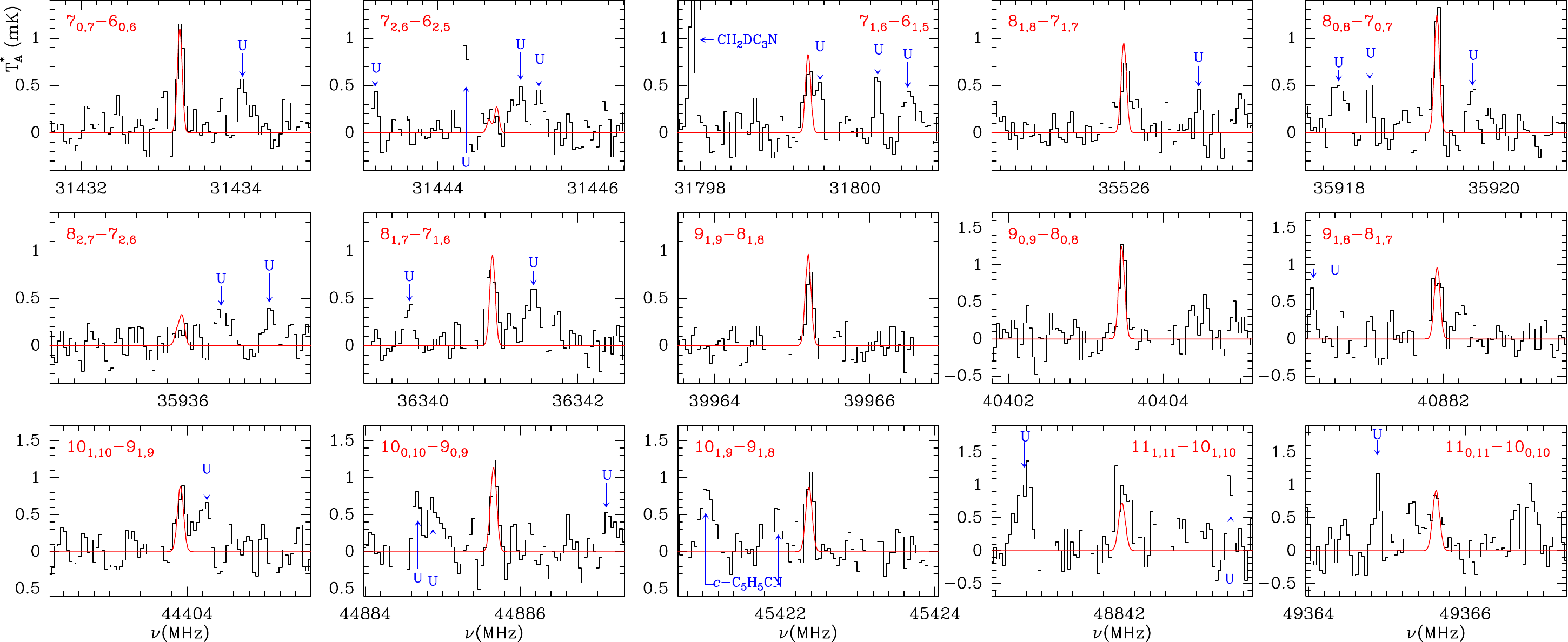}
\caption{Observed lines of $t$-CH$_3$CHCHCN ($trans$-crotononitrile) towards \mbox{TMC-1}.
Line parameters are given in Table \ref{line_par_t-ch3chchcn}.
The abscissa corresponds to the rest frequency assuming a local standard of rest velocity of 5.83
km s$^{-1}$. 
The ordinate is the antenna temperature corrected for atmospheric and telescope losses in milli Kelvin.
The red line shows the synthetic spectrum derived for
T$_{rot}$=7\,K and N($t$-CH$_3$CHCHCN)=5.0$\times$10$^{10}$ cm$^{-2}$.
Blank channels correspond to negative features produced
in the folding of the frequency switching data. 
The hyperfine structure has been included in the model.}
\label{fig_t-ch3chchcn}
\end{figure*}

\section{Results} \label{results}
QUIJOTE has now reached a level of sensitivity (0.15-0.4 mK across the Q-band) that permits new
isotopologues and CN and CCH derivatives of species with abundances $\sim$10$^{-9}$ to be detected.
Although it has not reached the confusion limit yet, special care has to be taken
when assigning lines to a given molecule as blending with other features often
occurs. Consequently, spectral line assignment has to be performed with special attention.
Line identification in this work was done using the catalogues 
MADEX \citep{Cernicharo2012}, CDMS \citep{Muller2005}, and JPL \citep{Pickett1998}. 
By May 2022, the MADEX code contained 6434 spectral
entries corresponding to the ground and vibrationally excited states, together
with the corresponding isotopologues, of 1734 molecules. 
Once the assignment of all known molecules and their isotopologues is done, then QUIJOTE permits one
to search for new molecular species for which laboratory spectroscopy is available. Last
but not least, QUIJOTE also allows
one to perform rotational spectroscopy in space of new species for which no 
previous rotational spectroscopic laboratory information was available such as HC$_5$NH$^+$ \citep{Marcelino2020}, 
HC$_3$O$^+$\citep{Cernicharo2020}, HC$_3$S$^+$ \citep{Cernicharo2021e}, CH$_3$CO$^+$ \citep{Cernicharo2021f},
HCCS$^+$ \citep{Cabezas2022a}, C$_5$H$^+$ \citep{Cernicharo2022}, HC$_7$NH$^+$ \citep{Cabezas2022b},
and HCCNCH$^+$ \citep{Agundez2022b}.

\subsection{Energetics of the cyano derivatives of propene}
All the molecular species studied in this work belong to the C$_4$H$_5$N family. Up 
to 16 isomers have been calculated within an energy window of 190 kJ mol$^{-1}$, where 
pyrrole, stabilized by aromaticity, is unambiguously the most stable species \citep{Lattelais2010}. 
According to \citet{Lattelais2010}, the second most stable species is crotononitrile lying at 33 kJ mol$^{-1}$
above pyrrole. Since \citet{Lattelais2010} did not distinguish between $cis$- and $trans$- crotononitrile, we ran 
our own energy calculations using the M{\o}ller-Plesset post-Hartree-Fock method \citep{Moller1934}(MP2) 
\citep{Moller1934} and the Pople basis set 6-311++G(d,p) \citep{Frisch1984}. Our results are similar to those 
found by \citet{Lattelais2010} with pyrrole being the most stable one and $trans$-crotononitrile lying at 44 
kJ mol$^{-1}$. The relative energies for the other five species are shown in Fig. \ref{fig_structures}. It must be noted that 
there exists another conformer for the ally cyanide isomer, named $trans$-allyl cyanide, which is about 
9 kJ mol$^{-1}$ less stable than the $gauche$ one. However, it has not been included in this work since 
no experimental data are available for it, contrary to the five species shown in Fig. \ref{fig_structures}
for which several laboratory investigations have been performed (see below). 
Moreover, taking
the weakness expected for its lines  into account, a search for it within the forest of U-lines of QUIJOTE is 
hopeless at this level of sensitivity.
Rotational constants for the five cyano derivatives studied in this work (shown in Fig. \ref{fig_structures}) 
were derived from a fit to the available laboratory spectroscopy and
have been implemented in MADEX \citep{Cernicharo2012} to predict the frequencies of their rotational
transitions and to compute their synthetic spectra.

\subsection{Detection of $trans$ and $cis$-crotononitrile ($t$/$c$-CH$_3$CHCHCN)}
Rotational spectra up to 371.1 GHz have been investigated for $cis$-crotononitrile and $trans$-crotononitrile
by \citet{Lesarri1995}. Their dipole moments have been measured by \citet{Beaudet1963} and 
\citet{Suzuki1970}, respectively. For $cis$-crotononitrile, internal rotation splittings were observed, but 
not for $trans$-crotononitrile \citep{Lesarri1995}.
The analysis of the hyperfine structure due to the $^{14}$N nuclear 
quadrupole has been done for $cis$-crotononitrile and $trans$-crotononitrile 
by \citet{Lesarri1995}, and very recently by \citet{McCarthy2020}. 

We started the search for both isomers of crotononitrile by adopting a rotational temperature of 10 K and
a column density of 10$^{11}$ cm$^{-2}$. All lines of the $trans$ and $cis$ isomers with $K_a\le2$ have been searched for.
Lines with $K_a\ge$3 are predicted to be extremely weak for the adopted rotational temperature. As a result of 
the inspection of the QUIJOTE data, all $K_a$=0,1
lines of the $trans$ isomer have been detected with no missing lines ($\mu_a$=4.39\,D, \citealt{Beaudet1963}). The
lines are shown in Fig. \ref{fig_t-ch3chchcn} and the line parameters are given in Table \ref{line_par_t-ch3chchcn}. 
Two lines with $K_a$=2 
are also presented to show the low intensity level expected for them
($7_{2,6}-6_{2,5}$ and $8_{2,7}-7_{2,6}$). In order to derive a column density and a rotational temperature
for this isomer, 
we have assumed a source of uniform brightness with a diameter of 80$''$ \citep{Fosse2001} and used a line model
fitting method (see, e.g. \citealt{Cernicharo2021e}). 
All models have been performed taking the hyperfine structure 
of the observed transitions into account. 
Although mostly unresolved, some line broadening could be produced in the lines 
of all isomers.
The derived rotational temperature is 7$\pm$1 K and the column 
density of $trans$-crotononitrile is (5.0$\pm$0.5)$\times$10$^{10}$
cm$^{-2}$. Adopting a column density for molecular hydrogen of 10$^{22}$ cm$^{-2}$ \citep{Cernicharo1987}, the
abundance of $trans$-crotononitrile is (5.0$\pm$0.5)$\times$10$^{-12}$. The detection of the $cis$-isomer is
reported in Appendix \ref{line_par_c-ch3chchcn}. For this isomer, we derived a rotational 
temperature of 7$\pm$1 K, identical to that of the $trans$ isomer, and a column density of (1.3$\pm$0.2)$\times$10$^{11}$, 
that is to say 2-3 times more abundant than $trans$-crotononitrile.

\subsection{Detection of methacrylonitrile (CH$_2$C(CH$_3$)CN)}
\begin{figure}
\centering
\includegraphics[width=0.48\textwidth]{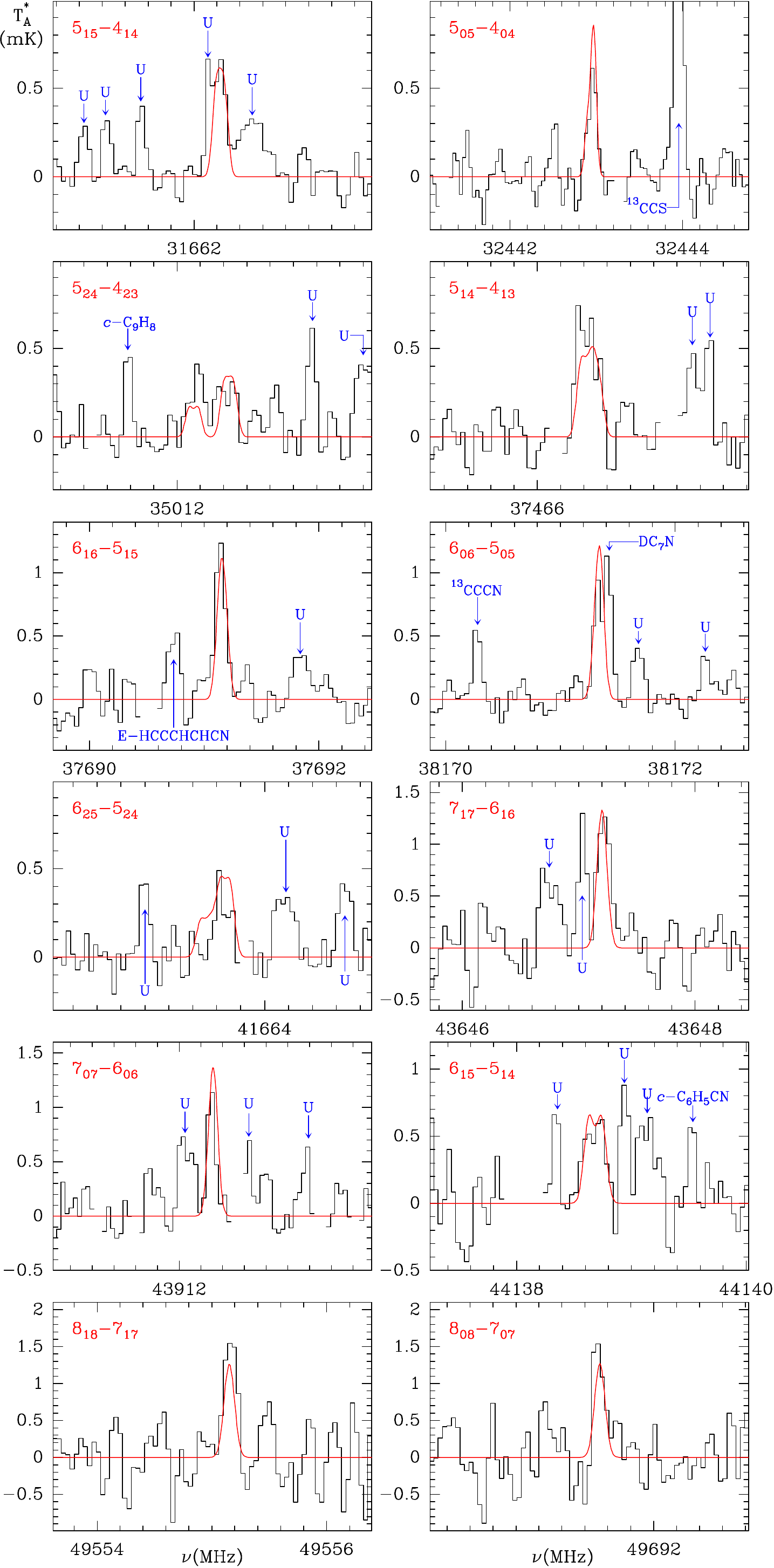}
\caption{Same as Fig.~\ref{fig_t-ch3chchcn}, but for CH$_2$C(CH$_3$)CN (methacrylonitrile). Line parameters are given in 
Table \ref{line_par-ch2cch3cn}. The red line shows the synthetic spectrum derived for
T$_{rot}$=7\,K and N(CH$_2$C(CN)CH$_3$)=1.0$\times$10$^{11}$ cm$^{-2}$.
The hyperfine structure has been included in the model.
}
\label{fig_ch2cch3cn}
\end{figure}

The rotational spectrum 
of methacrylonitrile has been investigated up to 200 GHz by \citet{Lopez1990}. This species shows small internal rotation 
splittings due to the methyl top. The dipole moments of this species
have been derived by Stark-effect measurements \citep{Lopez1990}. 
The analysis of the hyperfine structure due to the nuclear 
quadrupole coupling interactions of the $^{14}$N nucleus has been done by \citet{Lesarri1995}. All $K_a$=0,1 lines
have been detected as shown in Fig. \ref{fig_ch2cch3cn}. The line parameters are provided in Table \ref{line_par-ch2cch3cn}.
A couple of transitions with $K_a$=2 are also detected.
The derived rotational temperature is similar to that of crotononitrile, 7$\pm$1\,K, and the column density (total
$A$ + $E$) is (1.0$\pm$0.1)$\times$10$^{11}$ cm$^{-2}$. The abundance relative to H$_2$ is (1.0$\pm$0.1)$\times$10$^{-11}$.

\subsection{Detection of $gauche$- and $cis$-allyl cyanide ($g$/$c$-CH$_2$CHCH$_2$CN)}

\begin{figure}
\centering
\includegraphics[width=0.46\textwidth]{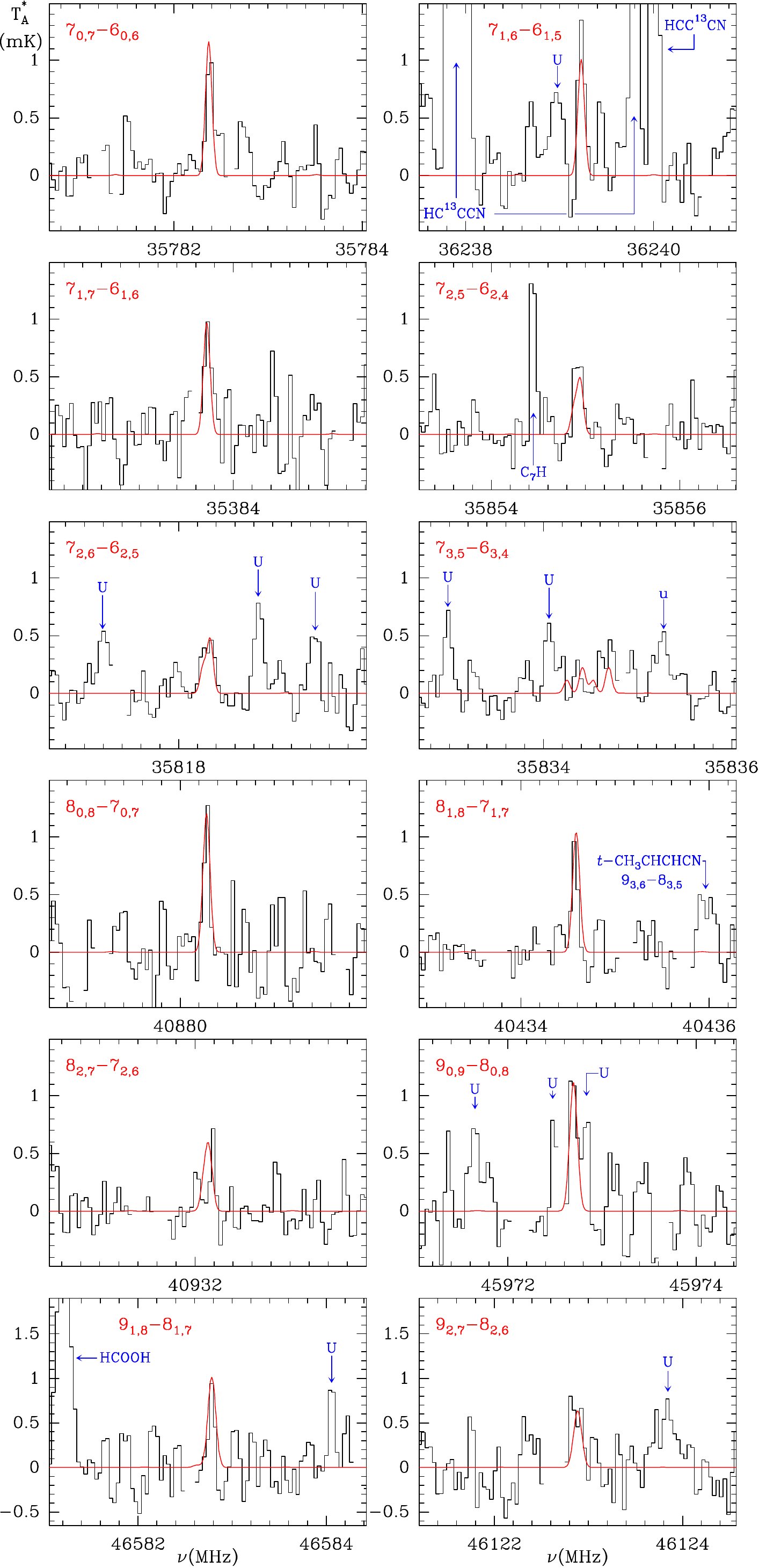}
\caption{Same as Fig.~\ref{fig_t-ch3chchcn}, but for $g$-CH$_2$CHCH$_2$CN ($gauche$-allyl cyanide).
Line parameters are given in Table \ref{line_par_g-ch2chch2cn}. 
The red line shows the synthetic spectrum derived for
T$_{rot}$=7\,K and N($g$-CH$_2$CHCH$_2$CN)=7.0$\times$10$^{10}$ cm$^{-2}$.
The hyperfine structure has been included in the model.
}
\label{fig_g-ch2chch2cn}
\end{figure}

The rotational spectra of the $cis$ and $gauche$ conformers of allyl cyanide have been measured up to 250 GHz by \citet{Demaison1991} 
and the dipole moments have been determined by \citet{Sastry1968}. 
The analysis of the hyperfine structure due to the nuclear 
quadrupole coupling interactions of the $^{14}$N nucleus has been done by \citet{McCarthy2020}. For 
the $gauche$ conformer, all $K_a$=0,1
lines are detected in the frequency coverage of the QUIJOTE line survey. The lines are shown in 
Fig. \ref{fig_g-ch2chch2cn}
and the line parameters are given in Table \ref{line_par_g-ch2chch2cn}. Similar to the case of the other isomers,
the lines with $K_a$=2,3 are weak and below the sensitivity of the survey. The best rotational temperature is
also 7$\pm$1\,K and the column density is (8.0$\pm$0.8)$\times$10$^{10}$ cm$^{-2}$.

The lines of $cis$-allyl cyanide ($c$-CH$_2$CHCH$_2$CN) are shown in Fig. \ref{fig_c-ch2chch2cn} and
the line parameters are given in Table \ref{line_par-c-ch2chch2cn}. All $K_a$=0,1 lines are detected,
together with two $K_a$=2 transitions. The best rotational temperature is also 7$\pm$1\,K and the column
density is N($c$-CH$_2$CHCH$_2$CN)=(7.0$\pm$0.7)$\times$10$^{10}$ cm$^{-2}$, that is to say nearly identical to
that of the $gauche$ conformer.

\section{Discussion}

The five isomers of cyano-propene detected in this work show very similar abundances, despite the large energy differences between them. This is a further example that the chemical composition 
of cold dark clouds is driven by chemical kinetics so that thermodynamic considerations such as the minimum energy 
principle \citep{Lattelais2010} are not relevant.

The formation of the five cyano derivatives of propene is likely related to propene, the formation of 
which is not yet well understood. Soon after the discovery of propene in \mbox{TMC-1} by \cite{Marcelino2007}, \cite{Herbst2010} 
proposed a formation mechanism based on two successive radiative associations of the propargyl ion (CH$_2$CCH$^+$) with 
H$_2$, resulting in the ion C$_3$H$_7^+$, the dissociative recombination of which would yield propene. This mechanism 
was indeed found to produce propene with abundances in excess of that observed in \mbox{TMC-1} (e.g. \citealt{Agundez2013}). 
However, further experiments and calculations have shown that the proposed radiative associations involving H$_2$ have activation 
barriers and thus cannot be efficient at the low temperatures of \mbox{TMC-1} \citep{Lin2013}. Alternative mechanisms involving 
dust grains have been proposed for the formation of propene in cold dark clouds \citep{Rawlings2013,Hickson2016,Abplanalp2018}. 
That is, currently, there is no gas-phase mechanism able to explain the formation of propene in dark clouds.

Regardless of pathway to propene in cold dark clouds, the large abundance in which it is present makes 
it feasible to form C$_2$H and CN derivatives since it has been found for other hydrocarbons such as C$_2$H$_4$ \citep{Cernicharo2021d}, 
CH$_3$CCH and CH$_2$CCH$_2$ \citep{Marcelino2021,Cernicharo2021b}, $c$-C$_3$H$_2$ \citep{Cernicharo2021c}, 
benzene \citep{McGuire2018}, and cyclopentadiene \citep{Cernicharo2021h}. In this line, the most obvious route to the cyano-propene isomers detected in 
\mbox{TMC-1} is the reaction between CN and propene. This reaction has been measured to be rapid at temperatures as low as 
23 K \citep{Morales2010} and the mechanism and products have been investigated by means of crossed beam experiments and 
theoretical calculations \citep{Gu2008,Huang2009}. We can consider the following reaction channels
\begin{subequations}
\begin{align}
\rm CN + CH_2CHCH_3 & \rightarrow \rm CH(CN)CHCH_3 + H, \\
                                        & \rightarrow \rm CH_2C(CN)CH_3 + H, \\
                                        & \rightarrow \rm CH_2CHCH_2(CN) + H, \\
                                        & \rightarrow \rm CH_2CHCN + CH_3,
\end{align}
\end{subequations}
where channels (1a), (1b), and (1c) produce crotononitrile, methacrylonitrile, and allyl cyanide, respectively, that is to say the three isomers 
of cyano-propene with the CN substituted at the three different carbon positions of propene (see Fig.~\ref{fig_structures}). 
\cite{Huang2009} found that formation of vinyl cyanide (channel 1d) dominates, with a branching ratio in the range 70-86\,\% at 
zero collision energy. Given that vinyl cyanide has been detected in \mbox{TMC-1} with a column density of 6.5$\times$10$^{12}$ cm$^{-2}$ 
\citep{Cernicharo2021d}, which is well above those derived here for the different cyano-propene isomers, this reaction could indeed produce CH$_2$CHCN as a major 
product in cold dark clouds. In regards to the three isomers of cyano-propene, \cite{Huang2009} found that crotononitrile and 
allyl cyanide are formed with similar yields while methacrylonitrile is not formed. On the other hand, our observations indicate 
that the three isomers are present with similar abundances in \mbox{TMC-1}. The study of \cite{Huang2009} and our observations 
agree in that crotononitrile and allyl cyanide are formed with comparable abundances, but they disagree regarding methacrylonitrile. 
It would be surprising if this isomer is formed by a reaction different to that forming the other cyano-propene isomers, which 
makes revisiting the product distribution of reaction (1) worth it.

We included reaction (1) with a rate coefficient of 3.7\,$\times$\,10$^{-10}$ (as measured 
at 23 K; \citealt{Morales2010}) and a branching ratio of 20\,\% for the formation of cyano-propene (as predicted by \citealt{Huang2009}) 
in a gas-phase chemical model similar to those presented in previous works (e.g. \citealt{Agundez2021}), and assumed that cyano-propene 
is mostly destroyed through reactions with C$^+$, HCO$^+$, H$_3^+$, H$^+$, and He$^+$. We found that the calculated abundance of 
cyano-propene is around 1\,\% of that or propene, which is in agreement with observations and supports reaction (1) as the main route 
to cyano-propene in \mbox{TMC-1}.

We searched for the two ethynyl derivatives of propene for which laboratory data are available from \citet{McCarthy2020},
$E$- and $Z$-3-penten-1-yne (CH$_3$CHCHCCH). 
The reaction of propene and CCH has been studied down to 79\,K and has been found to be rapid \citep{Bouwman2012}.
However, the main product of the reaction is vinyl acetylene with a yield of only 15\% for the isomers of C$_5$H$_6$. Moreover, the main
isomer of C$_5$H$_6$ produced in the reaction is 4-penten-1-yne, with a 3-penten-1-yne yield $\sim$2 times lower.
Only 3$\sigma$ upper limits of 5$\times$10$^{11}$ cm$^{-2}$ and of 9$\times$10$^{11}$ cm$^{-2}$ have been obtained for the $E$ and 
$Z$ isomers of 3-penten-1-yne, respectively.

\section{Conclusions}

We have reported the detection of five isomers of cyano-propene with very similar abundances towards the cold dark cloud \mbox{TMC-1}. 
The comparison of observations and models suggests that these cyano derivatives are produced in
the gas-phase by the reaction of cyanide with propene. Only upper limits are obtained for the two isomers of the 
ethynyl derivatives of propene, $E$- and $Z$-CH$_3$CHCHCCH.

\begin{acknowledgements}
We thank ERC for funding
through grant ERC-2013-Syg-610256-NANOCOSMOS. We also thank Ministerio de Ciencia e Innovaci\'on of Spain (MICIU) for funding support through projects
PID2019-106110GB-I00, PID2019-107115GB-C21 / AEI / 10.13039/501100011033, 
and PID2019-106235GB-I00. 

\end{acknowledgements}

\begin{appendix}
\section{Line parameters}
\label{line_parameters}
Line parameters for the different molecules studied in this work were obtained by fitting a Gaussian line
profile to the observed data. A window of $\pm$ 15 \kms\, around the v$_{LSR}$ of the source (5.83 km s$^{-1}$)
has been considered for each transition.

\subsection{$trans$-crotononitrile}
\label{lin_par_t-ch3chchcn}

The line identification for $trans$-crotononitrile was done using the frequency predictions based on the laboratory measurements 
reported by \citet{Lesarri1995}. In that work, the authors measured a total of 75 $a$-type rotational transitions with 
$J_{max}$=78 and $K_{a,max}$=19 in the 4-420 GHz frequency range. The spectroscopic constants derived by \citet{Lesarri1995} 
from the analysis of these data are shown in Table \ref{tcroto_constants}. We used these molecular constants to predict the 
rotational transitions of $trans$-crotononitrile in the Q-band. We observed a total of 15 transitions in the 31.0-50.4 
TMC-1 survey. We detected all transitions $J_u$=7 through $J_u$ = 11 with $K_a$=0,1, and 
marginally
two lines with $K_a$=2 ($7_{2,6}-6_{2,5}$, and $8_{2,7}-7_{2,6}$). 
The derived line parameters are given in Table 
\ref{line_par_t-ch3chchcn}. The lines are shown in Fig. \ref{fig_t-ch3chchcn}. None of the lines detected in our survey 
were observed in the laboratory by \citet{Lesarri1995}, so we included the astronomical and laboratory frequencies in a combined analysis using the programme SPFIT 
\citep{Pickett1991}. The derived spectroscopic constants, shown in Table \ref{tcroto_constants}, 
are very similar to those reported by \citet{Lesarri1995}, with the exception of the parameter $\Phi_{J}$, which was not 
determined in the laboratory work.

\begin{figure}
\centering
\includegraphics[width=0.477\textwidth]{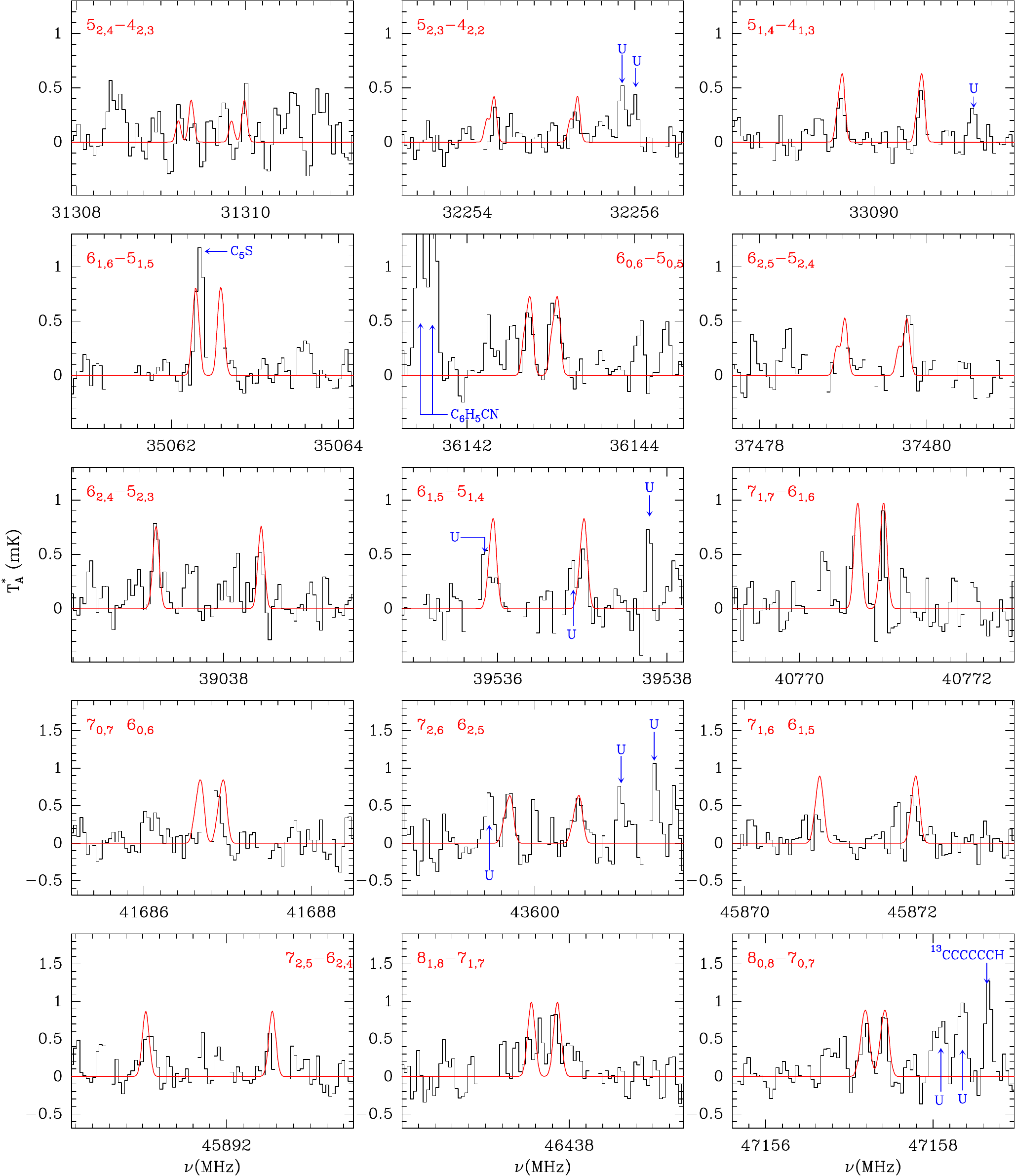}
\caption{Observed transitions of $cis$-CH$_3$CHCHCN ($cis$-crotononitrile) towards TMC-1.
Line parameters are given in Table \ref{line_par_c-ch3chchcn}. Internal rotation splits the
lines into two components, $E$ and $A$, which
correspond to the low and the high frequency components in each panel, respectively.
The abscissa corresponds to the rest frequency assuming a local standard of rest velocity of 5.83
km s$^{-1}$.
The ordinate is the antenna temperature corrected for atmospheric and telescope losses in milli Kelvin. The red line shows the synthetic spectrum derived for
T$_{rot}$=7\,K and N($cis$-CH$_3$CHCHCN)=1.3$\times$10$^{11}$ cm$^{-2}$.
}
\label{fig_c-ch3chchcn}
\end{figure}

\begin{table}
\small
\caption{Molecular constants MHz) for $trans$-crotononitrile.}
\label{tcroto_constants}
\centering
\begin{tabular}{{lcc}}
\hline
\hline
Constant          &       Lab$^a$+TMC-1           & Lab$^a$   \\
\hline
$A$              &       38053.422(36)$^b$        &      38053.406(43)           \\
$B$              &      2297.07475(42)            &     2297.06890(72)           \\
$C$              &      2195.19007(40)            &     2195.18358(60)           \\
$\Delta_{J}$     &       0.286585(143)10$^{-3}$   &      0.283422(46)10$^{-3}$   \\
$\Delta_{JK}$    &       -0.0173276(35)           &      -0.0173321(34)          \\
$\delta_{J}$     &     0.0297147(196)10$^{-3}$    &    0.029727(31)10$^{-3}$     \\
$\delta_{JK}$    &       -1.140(57)10$^{-3}$      &      -1.124(64)10$^{-3}$     \\
$\Phi_{J}$       &        0.2913(139)10$^{-9}$    &       -                      \\
$\Phi_{JK}$      &      -0.039316(217)10$^{-6}$   &     -0.039665(194)10$^{-6}$  \\
$\Phi_{KJ}$      &         1.8560(101)10$^{-6}$   &        1.8400(101)10$^{-6}$  \\
$rms$$^c$        &           41.8                 &          46.0                \\
N$^d$            &           90                   &          75                  \\
\hline
\end{tabular}
\tablefoot{
        \tablefoottext{a}{\citealt{Lesarri1995}.}\tablefoottext{b}{Values between parentheses correspond to the uncertainties of the parameters in units of the last 
		significant digits.} \tablefoottext{c}{The standard deviation of the fit in kHz.} \tablefoottext{d}{Number of lines included in the fit.}
    }
\end{table}
\normalsize

\begin{table*}
\caption{Observed line parameters for $trans$-CH$_3$CHCHCN.}
\label{line_par_t-ch3chchcn}
\centering
\begin{tabular}{{cccccc}}
\hline
{\textit Transition$^a$} & $\nu_{obs}^b$       & $\int$T$_A^*$dv $^c$     & $\Delta$v$^d$   & T$_A^*$& N\\
                     &  (MHz)              & (mK km\,s$^{-1}$)        & (km\,s$^{-1}$)  & (mK)   & \\
\hline
$ 7_{0, 7}- 6_{0, 6}$& 31433.290$\pm$0.010 & 1.02$\pm$0.10& 0.82$\pm$0.08& 1.17$\pm$0.13 & \\
$ 7_{2, 6}- 6_{2, 5}$& 31444.676$\pm$0.030 & 0.44$\pm$0.13& 1.56$\pm$0.40& 0.26$\pm$0.12 &A\\
$ 7_{1, 6}- 6_{1, 5}$& 31799.411$\pm$0.020 & 0.97$\pm$0.20& 1.43$\pm$0.33& 0.65$\pm$0.14 &B\\
$ 8_{1, 8}- 7_{1, 7}$& 35526.024$\pm$0.010 & 0.74$\pm$0.13& 0.90$\pm$0.19& 0.77$\pm$0.13 & \\
$ 8_{0, 8}- 7_{0, 7}$& 35919.271$\pm$0.010 & 1.19$\pm$0.10& 0.80$\pm$0.08& 1.40$\pm$0.13 & \\
$ 8_{2, 7}- 7_{2, 6}$& 35935.969$\pm$0.000 &              &              & $\le$0.42     &C\\
$ 8_{1, 7}- 7_{1, 6}$& 36340.885$\pm$0.010 & 1.01$\pm$0.07& 1.14$\pm$0.31& 0.83$\pm$0.11 & \\
$ 9_{1, 9}- 8_{1, 8}$& 39965.235$\pm$0.010 & 0.62$\pm$0.08& 0.73$\pm$0.12& 0.80$\pm$0.12 & \\
$ 9_{0, 9}- 8_{0, 8}$& 40403.479$\pm$0.010 & 1.01$\pm$0.13& 0.72$\pm$0.10& 1.32$\pm$0.17 & \\
$ 9_{1, 8}- 8_{1, 7}$& 40881.992$\pm$0.010 & 0.99$\pm$0.12& 1.05$\pm$0.13& 0.88$\pm$0.16 & \\
$10_{1,10}- 9_{1, 9}$& 44403.934$\pm$0.010 & 0.89$\pm$0.18& 1.06$\pm$0.26& 0.79$\pm$0.20 & \\
$10_{0,10}- 9_{0, 9}$& 44885.670$\pm$0.010 & 0.89$\pm$0.13& 0.69$\pm$0.11& 1.22$\pm$0.21 & \\
$10_{1, 9}- 9_{1, 8}$& 45422.385$\pm$0.010 & 0.82$\pm$0.10& 0.68$\pm$0.09& 1.13$\pm$0.18 & \\
$11_{1,11}-10_{1,10}$& 48842.074$\pm$0.020 & 1.16$\pm$0.26& 1.16$\pm$0.30& 0.94$\pm$0.26 &B\\
$11_{0,11}-10_{0,10}$& 49365.622$\pm$0.020 & 0.43$\pm$0.14& 0.42$\pm$0.18& 0.97$\pm$0.29 & \\
\hline
\end{tabular}
\tablefoot{\\
        \tablefoottext{a}{Quantum numbers are $J, K_a, K_c$.}\\
        \tablefoottext{b}{Observed frequencies adopting a v$_{LSR}$ of 5.83 km s$^{-1}$ for TMC-1.}\\
        \tablefoottext{c}{Integrated line intensity in mK km\,s$^{-1}$.}\\
        \tablefoottext{d}{Linewidth at half intensity derived by fitting a Gaussian line profile to the observed
     transitions (in km\,s$^{-1}$).}\\
\tablefoottext{A}{Marginally detected.}\\
\tablefoottext{B}{Partially blended with another feature.}\\
\tablefoottext{C}{3$\sigma$ upper limit. Frequency corresponds to the predicted one.}
}
\end{table*}
\normalsize

\subsection{$cis$-crotononitrile}
\label{lin_par_c-ch3chchcn}

Laboratory measurements of the rotational spectrum of $cis$-crotononitrile are reported by \citet{Lesarri1995}.  In this work 
the rotational transitions of $cis$-crotononitrile are observed as doublets ($A$ and $E$) due to methyl top internal rotation. 
The dipole moment of the molecule, 3.74\,D \citep{Beaudet1963}, is slightly smaller than that of the $trans$ isomer. 
\citet{Lesarri1995} reported the experimental molecular constants of the $A$ state, obtained in a single state fit, but they did not perform a fit for the $E$ state.\ As a result of which, only the observed frequency transitions were reported. In this situation, the frequency predictions 
for $cis$-crotononitrile in the Q-band are not very accurate. Hence, we first carried out a double state ($A$ and $E$) fit using the 
frequencies reported by \citet{Lesarri1995}. We predicted the rotational spectra for $A$ and $E$ in the Q-band frequency range 
and we were able to observe a total of 22 lines that were included in the spectral analysis together with those 
measured by \citet{Lesarri1995}. The derived line parameters are given in Table \ref{line_par_c-ch3chchcn} and the lines 
are shown in Fig. \ref{fig_c-ch3chchcn}. In this manner, our set of data contains 182 rotational transitions and the 
results of the fit done with the XIAM programme \citep{Hartwig1996} are shown in Table \ref{ccroto_constants}. The molecular constants are 
very similar to those reported by \citet{Lesarri1995}, even the geometrical parameters are also in agreement 
in both fits.

\begin{figure}
\centering
\includegraphics[width=0.477\textwidth]{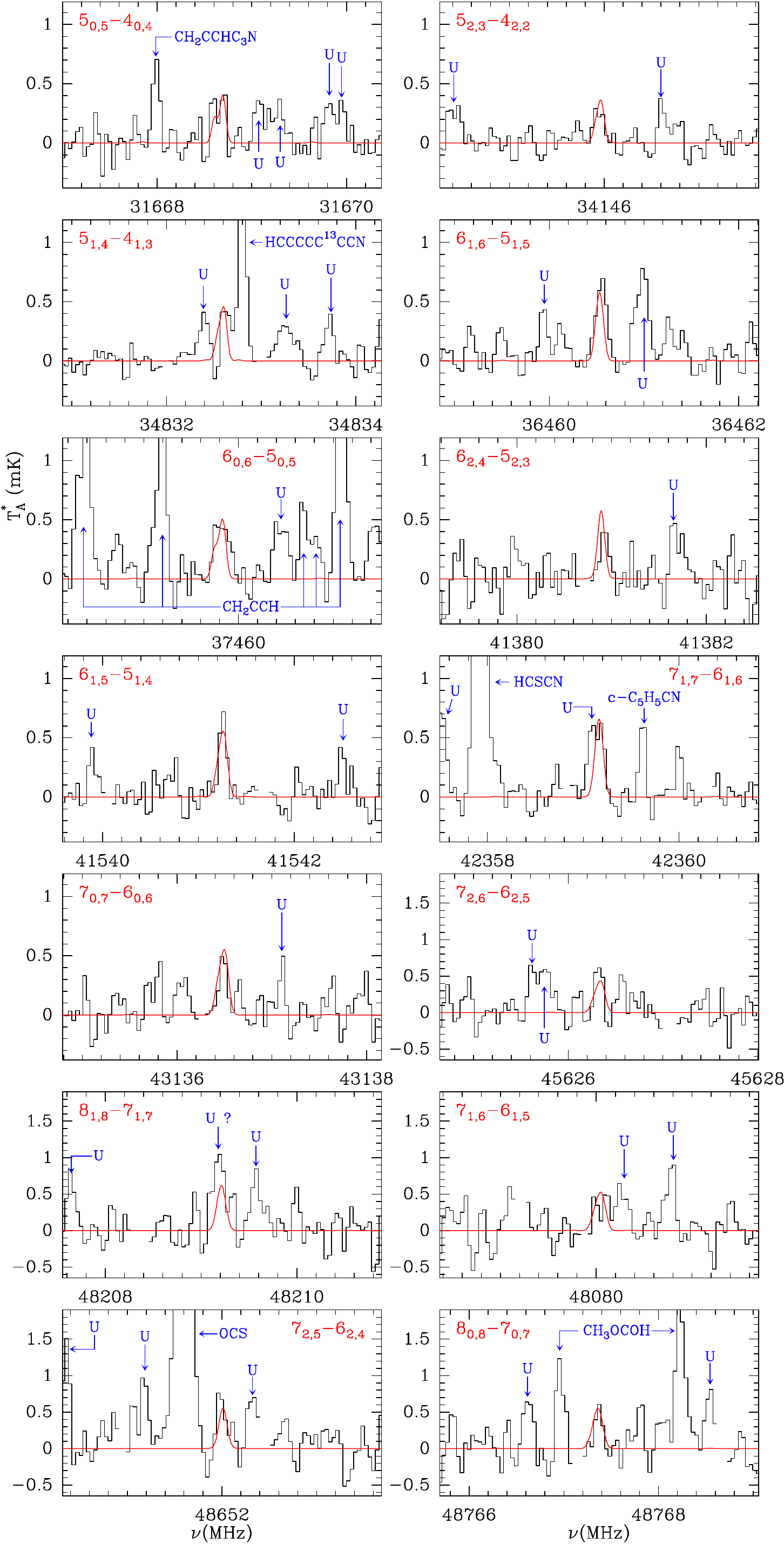}
\caption{Same as Fig.~\ref{fig_c-ch3chchcn}, but for $cis$-CH$_2$CHCH$_2$CN ($cis$-allyl cyanide).
Line parameters are given in Table \ref{line_par-c-ch2chch2cn}. The red line shows the synthetic spectrum derived for
T$_{rot}$=7\,K and N($cis$-CH$_2$CHCH$_2$CN)=7$\times$10$^{10}$ cm$^{-2}$.
The hyperfine structure has been included in the model.
}
\label{fig_c-ch2chch2cn}
\end{figure}

Contrarily to the $trans$ isomer, some of the expected strong lines of the $cis$ form are missing. Although in some cases this non-detection can be attributed 
to the presence of negative features produced in the folding of the frequency switching data, there is
a case for which an additional explanation is needed. It corresponds to the $E$ component of the $6_{1,6}-6_{1,5}$ transition predicted at 35062.586$\pm$0.003 MHz. 
The frequency switching data with a throw of 8 MHz produce an exact coincidence with one of the hyperfine components of the $8_{1,8}-7_{1,7}$ transition at  
35054.54 MHz of H$_2$C$_4$N, a radical recently detected with QUIJOTE \citep{Cabezas2021}. For the data with a frequency switching throw of 10 MHz, there is a 
U feature at 35072.559 MHz, that is, exactly at 10 MHz from the frequency of the line of $cis$-crotononitrile. Hence, both sets of data produce a negative feature 
at the frequency of the line, which explains the lack of observed emission. Furthermore, the $A$ component of this transition of $cis$-crotononitrile
is partially blended with the $J$=19-18 transition of C$_5$S for which an intensity of $\sim$1 mK has been previously reported towards TMC-1 
\citep{Cernicharo2021g}. Hence, we consider that there are enough individual lines  for a definitive detection of the $cis$ isomer with the present 
data. A fit to the data of Fig. \ref{fig_c-ch3chchcn} provides T$_{rot}$=8$\pm$1 K and N($cis$-CH$_3$CHCHCN)=(1.5$\pm$0.2)$\times$10$^{11}$ cm$^{-2}$, 
corresponding to an abundance of (1.5$\pm$0.2)$\times$10$^{-11}$.

\begin{table}
\small
\caption{Molecular constants (all in MHz except for those with specified units, see Notes) for $cis$-crotononitrile.}
\label{ccroto_constants}
\centering
\begin{tabular}{{lcc}}
\hline
\hline
Constant          &       Merged Fit$^a$           & Lab$^b$   \\
\hline
$A$                 &      11854.3814(55)$^c$              &     11854.4494(100)               \\
$B$                 &       3524.63261(92)                 &      3524.6977(145)               \\
$C$                 &       2759.72005(96)                 &      2759.74293(73)               \\
$\Delta_{J}$        &           2.72760(43)10$^{-3}$       &          2.73153(84)10$^{-3}$     \\
$\Delta_{JK}$       &         -18.2134(61)10$^{-3}$        &        -18.2106(53)10$^{-3}$      \\
$\Delta_{K}$        &          65.737(40)10$^{-3}$         &         65.801(59)10$^{-3}$       \\
$\delta_{J}$        &           0.87233(25)10$^{-3}$       &          0.87471(54)10$^{-3}$     \\
$\delta_{JK}$       &           5.222(16)10$^{-3}$         &          5.2526(167)10$^{-3}$     \\
$\Phi_{J}$          &         0.0051790(74)10$^{-6}$       &        0.005204(93)10$^{-6}$      \\
$\Phi_{KJ}$         &      -0.201(23)10$^{-6}$             &     -0.232(20)10$^{-6}$           \\
$\phi_{J}$          &         0.0022233(62)10$^{-6}$       &        0.002287(82)10$^{-6}$      \\
$V_3$$^d$           &               1373(17)               &              1387.2(7)            \\
$\angle$(a,$i$)$^e$ &               103.61(25)             &              102.751              \\
$I$$_{\alpha}$$^f$  &               3.141(82)              &              3.106                \\
$\rho$$^g$          &          0.02745(54)                 &         0.02655                   \\
$\beta$$^h$         &          2.2541(93)                  &         2.2213                    \\
$rms$$^i$           &           24.6                       &          34$^j$/51$^k$                    \\
N$^l$               &           182                        &          111$^j$/144$^k$                  \\
\hline
\end{tabular}
\tablefoot{
\tablefoottext{a}{Fit to the lines observed by \citealt{Lesarri1995}, and in this work towards TMC-1.}
\tablefoottext{b}{Molecular constants derived by \citet{Lesarri1995} from an $A$-state analysis.}\tablefoottext{c}{Values between parentheses correspond to the uncertainties of the parameters in units of the last significant digits.} \tablefoottext{d}{Internal rotation barrier height, in cal/mol.} \tablefoottext{e}{Angles between the methyl top axis and the inertial $a$-axis, in degrees.}\tablefoottext{f}{Moment of inertia of methyl top, in  u{$\AA$}$^2$.}\tablefoottext{g}{Internal-overall molecular rotation constant, no units.}\tablefoottext{h}{Angle between the rho system and the principal axes system, 
in rad.}\tablefoottext{i}{The standard deviation of the fit in kHz.} \tablefoottext{j}{From the $A$-state fitting analysis.}\tablefoottext{k}{From the $A$ and $E$ states' internal rotation analysis.}\tablefoottext{l}{Number of lines included in the fit.}
        
    }
\end{table}
\normalsize

\begin{table*}
\caption{Observed line parameters for $cis$-CH$_3$CHCHCN.}
\label{line_par_c-ch3chchcn}
\centering
\begin{tabular}{{cccccc}}
\hline
{\textit Transition$^a$} & $\nu_{obs}^b$       & $\int$T$_A^*$dv $^c$     & $\Delta$v$^d$   & T$_A^*$& N\\
                     &  (MHz)              & (mK km\,s$^{-1}$)        & (km\,s$^{-1}$)  & (mK)   & \\
\hline
$5_{2,4}-4_{2,3}$ $E$& 31309.283$\pm$0.032&                        &              &<0.33&A\\
$5_{2,4}-4_{2,3}$ $A$& 31309.939$\pm$0.002&                        &              &<0.33&A\\
$5_{2,3}-4_{2,2}$ $E$& 32254.342$\pm$0.015& 0.21$\pm$0.06& 0.67$\pm$0.18& 0.36$\pm$0.10& \\ 
$5_{2,3}-4_{2,2}$ $A$& 32255.294$\pm$0.015& 0.28$\pm$0.07& 0.84$\pm$0.22& 0.32$\pm$0.10& \\ 
$5_{1,5}-4_{1,3}$ $E$& 33089.608$\pm$0.010& 0.38$\pm$0.08& 0.87$\pm$0.10& 0.42$\pm$0.l0& \\ 
$5_{1,5}-4_{1,3}$ $A$& 33090.574$\pm$0.010& 0.41$\pm$0.08& 0.75$\pm$0.19& 0.51$\pm$0.l0& \\ 
$6_{1,6}-5_{1,5}$ $E$& 35062.310$\pm$0.047&                        &              &    &B\\
$6_{1,6}-5_{1,5}$ $A$& 35062.586$\pm$0.003&                        &              &    &C\\
$6_{0,6}-5_{0,5}$ $E$& 36142.741$\pm$0.010& 0.58$\pm$0.08& 0.87$\pm$0.13& 0.62$\pm$0.10& \\
$6_{0,6}-5_{0,5}$ $A$& 36143.051$\pm$0.010& 0.90$\pm$0.10& 1.29$\pm$0.16& 0.66$\pm$0.10& \\
$6_{2,5}-5_{2,4}$ $E$& 37478.991$\pm$0.011&                        &              &    &D\\
$6_{2,5}-5_{2,4}$ $A$& 37479.785$\pm$0.010& 0.47$\pm$0.10& 0.71$\pm$0.15& 0.62$\pm$0.12& \\
$6_{2,4}-5_{2,3}$ $E$& 39037.170$\pm$0.010& 0.41$\pm$0.07& 0.68$\pm$0.13& 0.56$\pm$0.10& \\ 
$6_{2,4}-5_{2,3}$ $A$& 39038.454$\pm$0.010& 0.51$\pm$0.08& 0.90$\pm$0.18& 0.53$\pm$0.10& \\ 
$6_{1,5}-5_{1,4}$ $E$& 39535.973$\pm$0.010& 0.29$\pm$0.12& 0.92$\pm$0.20& 0.29$\pm$0.14& \\
$6_{1,5}-5_{1,4}$ $A$& 39537.017$\pm$0.010& 0.38$\pm$0.09& 0.59$\pm$0.16& 0.60$\pm$0.14& \\
$7_{1,7}-6_{1,6}$ $E$& 40770.659$\pm$0.020& 0.30$\pm$0.10& 0.42$\pm$0.20& 0.47$\pm$0.14& \\ 
$7_{1,7}-6_{1,6}$ $A$& 40770.994$\pm$0.010& 0.40$\pm$0.09& 0.53$\pm$0.12& 0.78$\pm$0.14& \\ 
$7_{0,7}-6_{0,6}$ $E$& 41686.672$\pm$0.020& 0.27$\pm$0.06& 0.48$\pm$0.12& 0.53$\pm$0.11& \\ 
$7_{0,7}-6_{0,6}$ $A$& 41686.899$\pm$0.015& 0.67$\pm$0.10& 0.96$\pm$0.13& 0.66$\pm$0.11& \\ 
$7_{2,6}-6_{2,5}$ $E$& 43599.700$\pm$0.010& 0.58$\pm$0.12& 0.73$\pm$0.18& 0.75$\pm$0.24& \\
$7_{2,6}-6_{2,5}$ $A$& 43600.526$\pm$0.010& 0.66$\pm$0.15& 1.18$\pm$0.29& 0.52$\pm$0.24& \\
$7_{1,6}-6_{1,5}$ $E$& 45870.873$\pm$0.015& 0.32$\pm$0.08& 0.53$\pm$0.14& 0.57$\pm$0.14& \\ 
$7_{1,6}-6_{1,5}$ $A$& 45872.008$\pm$0.015& 0.46$\pm$0.08& 0.65$\pm$0.12& 0.67$\pm$0.14& \\ 
$7_{2,5}-6_{2,4}$ $E$& 45891.080$\pm$0.020& 0.61$\pm$0.14& 0.38$\pm$0.17& 0.57$\pm$0.19& \\ 
$7_{2,5}-6_{2,4}$ $A$& 45892.511$\pm$0.050& 0.26$\pm$0.09& 0.36$\pm$0.17& 0.63$\pm$0.19&E\\ 
$8_{1,8}-7_{1,7}$ $E$& 46437.647$\pm$0.020& 0.40$\pm$0.10& 0.48$\pm$0.13& 0.78$\pm$0.19& \\ 
$8_{1,8}-7_{1,7}$ $A$& 46437.819$\pm$0.020& 0.54$\pm$0.14& 0.59$\pm$0.12& 0.87$\pm$0.19& \\ 
$8_{0,8}-7_{0,7}$ $E$& 47157.196$\pm$0.015& 0.54$\pm$0.10& 0.80$\pm$0.16& 0.64$\pm$0.16& \\ 
$8_{0,8}-7_{0,7}$ $A$& 47157.416$\pm$0.015& 0.51$\pm$0.09& 0.55$\pm$0.10& 0.91$\pm$0.16& \\ 
\hline
\end{tabular}
\tablefoot{\\
        \tablefoottext{a}{Quantum numbers are $J, K_a, K_c$.}\\
        \tablefoottext{b}{Observed frequencies adopting a v$_{LSR}$ of 5.83 km s$^{-1}$ for TMC-1.}\\
        \tablefoottext{c}{Integrated line intensity in mK km\,s$^{-1}$.}\\
        \tablefoottext{d}{Linewidth at half intensity derived by fitting a Gaussian line profile to the observed
     transitions (in km\,s$^{-1}$).}\\
\tablefoottext{A}{3$\sigma$ upper limit. Frequency corresponds to the predicted one.}\\
\tablefoottext{B}{Heavily blended with the $J$=19-18 transition of C$_5$S.}\\
\tablefoottext{C}{In both set of data (8 and 10 MHz frequency switching throw), the line is blended with other features (see text).}\\
\tablefoottext{D}{Affected by a negative feature in the folding of the frequency  switching data. Frequency corresponds to the
predicted one.}
\tablefoottext{e}{Affected by a negative feature in the folding of the frequency  switching data.
 Line parameters are very uncertain.}
}
\end{table*}
\normalsize

\subsection{Methacrylonitrile, CH$_2$C(CH$_3$)CN}
\label{lin_par_ch2cch3cn}

The rotational spectrum of methacrylonitrile has been investigated in the laboratory by \citet{Lopez1990} and the 
analysis of the hyperfine structure due to the quadrupole of the $^{14}$N nucleus has been 
done by \citet{Lesarri1995}. Hence the frequency predictions for the $A$ and $E$ states including the nuclear 
quadrupole hyperfine structure have been done using those experimental data. Since the methyl internal rotation 
splittings are not large for methacrylonitrile, the data reported by \citet{Lopez1990} provide accurate 
predictions in the Q-band, as can be seen in Fig. \ref{fig_ch2cch3cn}. The 
line parameters are provided in Table \ref{line_par-ch2cch3cn}.

\begin{table*}
\caption{Observed line parameters for CH2C(CH3)CN (methacrylonitrile).}
\label{line_par-ch2cch3cn}
\centering
\begin{tabular}{{cccccc}}
\hline
{\textit Transition$^a$} & $\nu_{obs}^b$       & $\int$T$_A^*$dv $^c$     & $\Delta$v$^d$   & T$_A^*$& N\\
                     &  (MHz)              & (mK km\,s$^{-1}$)        & (km\,s$^{-1}$)  & (mK)   & \\
\hline
5$_{1,5}$-4$_{1,4}$   &  31662.232$\pm$0.015 &     0.78$\pm$0.21 &   1.11$\pm$ 0.36 &     0.66$\pm$0.07 &A\\
5$_{0,5}$-4$_{0,4}$   &  32442.960$\pm$0.010 &     0.66$\pm$0.10 &   1.07$\pm$ 0.19 &     0.58$\pm$0.11 & \\
5$_{2,4}$-4$_{2,3}$ E &  35012.374$\pm$0.020 &     0.36$\pm$0.11 &   1.14$\pm$ 0.45 &     0.30$\pm$0.08 & \\
5$_{2,4}$-4$_{2,3}$ A &  35012.479$\pm$0.020 &     0.25$\pm$0.10 &   0.69$\pm$ 0.28 &     0.34$\pm$0.08 & \\
5$_{1,4}$-4$_{1,3}$ E &  37466.359$\pm$0.010 &     0.51$\pm$0.13 &   0.63$\pm$ 0.16 &     0.77$\pm$0.11 &B\\
5$_{1,4}$-4$_{1,3}$ A &  37466.461$\pm$0.010 &     0.47$\pm$0.14 &   0.66$\pm$ 0.21 &     0.67$\pm$0.11 &B\\
6$_{1,6}$-5$_{1,5}$   &  37691.139$\pm$0.010 &     1.14$\pm$0.14 &   0.88$\pm$ 0.14 &     1.22$\pm$0.14 & \\
6$_{0,6}$-5$_{0,5}$   &  38171.329$\pm$0.010 &     0.73$\pm$0.10 &   0.72$\pm$ 0.12 &     0.95$\pm$0.09 &C\\
6$_{2,5}$-5$_{2,4}$ E &  41663.609$\pm$0.030 &     0.16$\pm$0.07 &   0.35$\pm$ 0.16 &     0.45$\pm$0.09 &B\\
6$_{2,5}$-5$_{2,4}$ A &  41663.689$\pm$0.030 &     0.24$\pm$0.10 &   0.86$\pm$ 0.29 &     0.26$\pm$0.09 &B\\
7$_{1,7}$-6$_{1,6}$   &  43647.218$\pm$0.010 &     1.26$\pm$0.22 &   0.90$\pm$ 0.19 &     1.32$\pm$0.17 & \\
7$_{0,7}$-6$_{0,6}$   &  43912.294$\pm$0.010 &     0.74$\pm$0.07 &   0.57$\pm$ 0.07 &     1.21$\pm$0.15 &D\\
6$_{1,5}$-5$_{1,4}$ E &  44138.607$\pm$0.010 &     0.42$\pm$0.05 &   0.79$\pm$ 0.09 &     0.50$\pm$0.16 &B\\
6$_{1,5}$-5$_{1,4}$ A &  44138.734$\pm$0.010 &     0.49$\pm$0.06 &   0.75$\pm$ 0.08 &     0.62$\pm$0.22 &B\\
8$_{1,8}$-7$_{1,7}$   &  49555.161$\pm$0.010 &     1.36$\pm$0.23 &   0.74$\pm$ 0.13 &     1.73$\pm$0.23 & \\
8$_{0,8}$-7$_{0,7}$   &  49691.519$\pm$0.010 &     1.10$\pm$0.20 &   0.62$\pm$ 0.13 &     1.66$\pm$0.20 & \\                                   
\hline
\end{tabular}
\tablefoot{\\
        \tablefoottext{a}{Quantum numbers are $J, K_a, K_c$.}\\
        \tablefoottext{b}{Observed frequencies adopting a v$_{LSR}$ of 5.83 km s$^{-1}$ for TMC-1.}\\
        \tablefoottext{c}{Integrated line intensity in mK km\,s$^{-1}$.}\\
        \tablefoottext{d}{Linewidth at half intensity derived by fitting a Gaussian line profile to the observed
     transitions (in km\,s$^{-1}$).}\\
\tablefoottext{A}{Line slightly blended with an unidentified line. Fit still possible and reliable.}\\
\tablefoottext{B}{$A$ and $E$ components are marginally resolved. Fit uncertain.}\\
\tablefoottext{C}{Line slightly blended with the $J$=35-34 line of DC$_7$N. Fit still possible and reliable.}\\
\tablefoottext{D}{Data set corresponds to frequency switching with a throw of 8 MHz alone.}\\
}
\end{table*}
\normalsize

\subsection{$gauche$-allyl cyanide, g-CH$_2$CHCH$_2$CN}
\label{lin_par_g-ch2chch2cn}
The derived line parameters for $gauche$-allyl cyanide ($g$-CH$_2$CHCH$_2$CN) are given in Table \ref{line_par_g-ch2chch2cn}.
The lines are shown in Fig. \ref{fig_g-ch2chch2cn}.

\begin{table*}
\caption{Observed line parameters for $gauche$-allyl cyanide.}
\label{line_par_g-ch2chch2cn}
\centering
\begin{tabular}{{cccccc}}
\hline
{\textit Transition$^a$} & $\nu_{obs}^b$       & $\int$T$_A^*$dv $^c$     & $\Delta$v$^d$   & T$_A^*$& N\\
                     &  (MHz)              & (mK km\,s$^{-1}$)        & (km\,s$^{-1}$)  & (mK)   & \\
\hline
7$_{1,7}$-6$_{1,6}$ &  35383.728$\pm$0.008 &     0.68$\pm$0.14 &   0.72$\pm$ 0.17 &     0.89$\pm$0.15 &A\\
7$_{0,7}$-6$_{0,6}$ &  35782.387$\pm$0.009 &     0.88$\pm$0.17 &   0.88$\pm$ 0.25 &     0.94$\pm$0.12 & \\
7$_{2,6}$-6$_{2,5}$ &  35818.345$\pm$0.014 &     0.38$\pm$0.11 &   0.78$\pm$ 0.26 &     0.45$\pm$0.12 &B\\
7$_{2,5}$-6$_{2,4}$ &  35854.937$\pm$0.010 &     0.50$\pm$0.13 &   0.71$\pm$ 0.22 &     0.66$\pm$0.14 & \\
7$_{3,5}$-6$_{3,4}$ &  35834.363$\pm$0.008 &                   &                  &       $\le$0.45   &C\\
7$_{3,4}$-6$_{3,3}$ &  35834.640$\pm$0.008 &                   &                  &       $\le$0.45   &C\\
7$_{1,6}$-6$_{1,5}$ &  36239.232$\pm$0.008 &     0.92$\pm$0.19 &   0.64$\pm$ 0.13 &     1.34$\pm$0.16 & \\
8$_{1,8}$-7$_{1,7}$ &  40434.570$\pm$0.007 &     0.51$\pm$0.10 &   0.51$\pm$ 0.10 &     0.95$\pm$0.13 & \\
8$_{0,8}$-7$_{0,7}$ &  40880.282$\pm$0.007 &     0.66$\pm$0.15 &   0.46$\pm$ 0.14 &     1.34$\pm$0.20 &A\\
8$_{2,7}$-7$_{2,6}$ &  40932.187$\pm$0.009 &     0.33$\pm$0.09 &   0.41$\pm$ 0.11 &     0.75$\pm$0.12 & \\
9$_{0,9}$-8$_{0,8}$ &  45972.696$\pm$0.008 &     0.83$\pm$0.13 &   0.66$\pm$ 0.13 &     1.19$\pm$0.15 &B\\
9$_{2,7}$-8$_{2,6}$ &  46122.881$\pm$0.023 &     0.56$\pm$0.19 &   0.85$\pm$ 0.31 &     0.62$\pm$0.19 & \\
9$_{1,8}$-8$_{1,7}$ &  46582.782$\pm$0.009 &     0.44$\pm$0.11 &   0.42$\pm$ 0.11 &     0.99$\pm$0.18 & \\
\hline
\end{tabular}
\tablefoot{\\
        \tablefoottext{a}{Quantum numbers are $J, K_a, K_c$.}\\
        \tablefoottext{b}{Observed frequencies adopting a v$_{LSR}$ of 5.83 km s$^{-1}$ for TMC-1.}\\
        \tablefoottext{c}{Integrated line intensity in mK km\,s$^{-1}$.}\\
        \tablefoottext{d}{Linewidth at half intensity derived by fitting a Gaussian line profile to the observed
     transitions (in km\,s$^{-1}$).}\\
\tablefoottext{A}{Frequency switching data with a throw of 10 MHz only.}\\
\tablefoottext{B}{Frequency switching data with a throw of 8 MHz only.}\\
\tablefoottext{C}{3$\sigma$ upper limit. Frequency corresponds to the predicted one.}
}
\end{table*}
\normalsize

\subsection{$cis$-allyl cyanide, c-CH$_2$CHCH$_2$CN}
\label{lin_par_c-ch2chch2cn}
The derived line parameters for $cis$-allyl cyanide ($c$-CH$_2$CHCH$_2$CN) are given in Table \ref{line_par-c-ch2chch2cn}.
The lines are shown in Fig. \ref{fig_c-ch2chch2cn}.

\begin{table*}
\caption{Observed line parameters for $cis$-allyl cyanide.}
\label{line_par-c-ch2chch2cn}
\centering
\begin{tabular}{{cccccc}}
\hline
{\textit Transition$^a$} & $\nu_{obs}^b$       & $\int$T$_A^*$dv $^c$     & $\Delta$v$^d$   & T$_A^*$& N\\
                     &  (MHz)              & (mK km\,s$^{-1}$)        & (km\,s$^{-1}$)  & (mK)   & \\
\hline
5$_{0,5}$-4$_{0,4}$ & 31668.651$\pm$0.030& 0.56$\pm$0.10& 1.43$\pm$0.23&0.37$\pm$0.10&A\\
5$_{2,4}$-4$_{2,3}$ & 32828.022$\pm$0.005&                        &                &$\le$0.30    &B\\
5$_{2,3}$-4$_{2,2}$ & 34415.938$\pm$0.020& 0.34$\pm$0.08& 1.11$\pm$0.28&0.28$\pm$0.09& \\
5$_{1,4}$-4$_{1,3}$ & 34832.525$\pm$0.030& 0.49$\pm$0.07& 0.96$\pm$0.14&0.48$\pm$0.09&C\\
6$_{1,5}$-5$_{1,5}$ & 36460.551$\pm$0.010& 0.82$\pm$0.11& 1.11$\pm$0.18&0.69$\pm$0.10&\\
6$_{0,6}$-5$_{0,5}$ & 37459.820$\pm$0.030& 0.78$\pm$0.10& 1.49$\pm$0.19&0.49$\pm$0.13&A,D\\
6$_{2,4}$-5$_{2,3}$ & 41380.929$\pm$0.030& 0.34$\pm$0.10& 0.73$\pm$0.25&0.43$\pm$0.15&\\
6$_{1,5}$-5$_{1,4}$ & 41541.255$\pm$0.010& 0.48$\pm$0.08& 0.60$\pm$0.11&0.75$\pm$0.13&\\
7$_{1,7}$-6$_{1,6}$ & 42359.196$\pm$0.020& 0.32$\pm$0.07& 0.52$\pm$0.18&0.57$\pm$0.11&E\\
7$_{0,7}$-6$_{0,6}$ & 43136.488$\pm$0.020& 0.36$\pm$0.09& 0.66$\pm$0.21&0.51$\pm$0.13&\\
7$_{2,6}$-6$_{2,5}$ & 45626.321$\pm$0.020& 0.43$\pm$0.09& 0.57$\pm$0.12&0.71$\pm$0.16&\\
7$_{3,4}$-6$_{3,3}$ & 47016.713$\pm$0.006&                        &                &$\le$0.54    &B\\
7$_{1,6}$-6$_{1,5}$ & 48080.066$\pm$0.030& 0.54$\pm$0.24& 1.30$\pm$0.60&0.39$\pm$0.20&F\\
8$_{1,8}$-7$_{1,7}$ & 48209.232$\pm$0.030& 0.66$\pm$0.15& 0.73$\pm$0.15&0.98$\pm$0.21&G\\
7$_{2,5}$-6$_{2,4}$ & 48651.988$\pm$0.020& 0.52$\pm$0.13& 0.62$\pm$0.18&0.79$\pm$0.22&\\
8$_{0,8}$-7$_{0,7}$ & 48767.369$\pm$0.020& 0.61$\pm$0.11& 0.49$\pm$0.18&0.63$\pm$0.22&\\

\hline
\end{tabular}
\tablefoot{\\
        \tablefoottext{a}{Quantum numbers are $J, K_a, K_c$.}\\
        \tablefoottext{b}{Observed frequencies adopting a v$_{LSR}$ of 5.83 km s$^{-1}$ for TMC-1.}\\
        \tablefoottext{c}{Integrated line intensity in mK km\,s$^{-1}$.}\\
        \tablefoottext{d}{Linewidth at half intensity derived by fitting a Gaussian line profile to the observed
     transitions (in km\,s$^{-1}$).}\\
\tablefoottext{A}{Line too broad. Hyperfine structure marginally resolved. A single line has been fitted.}\\
\tablefoottext{B}{3$\sigma$ upper limit. Frequency corresponds to the predicted one.}\\
\tablefoottext{C}{The line is possibly affected by a weak negative feature. Fit still possible, but the frequency and intensity are uncertain (see Fig.\ref{fig_c-ch2chch2cn}).}\\
\tablefoottext{D}{Frequency switching data with a throw of 8 MHz only.}\\
\tablefoottext{E}{Blended with an unknown feature. Fit still possible.}\\
\tablefoottext{F}{Marginal detection.}\\
\tablefoottext{G}{Line too strong, probably blended with a U line. Fit uncertain.}\\

}
\end{table*}
\normalsize

\end{appendix}
\end{document}